\documentclass{article}
\usepackage{spconf,amsmath,graphicx}
\usepackage{color,soul}
\usepackage{amssymb}
\usepackage{enumitem}
\setlist{nosep, leftmargin=14pt}

\usepackage{mwe} % to get dummy images

\title{Full-scale Deeply Supervised Attention Network for Segmenting COVID-19 Lesions}
\name
  {Pallabi Dutta\sthanks{Corresponding author is Pallabi Dutta (duttapallabi2907@gmail.com)}, Sushmita Mitra}
\address{Machine Intelligence Unit,
Indian Statistical Institute, Kolkata 700108, INDIA}
\begin{document}
%\ninept
%
\maketitle
\begin{abstract}
Automated delineation of COVID-19 lesions from lung CT scans aids the diagnosis and prognosis for patients. The asymmetric shapes and positioning of the infected regions make the task extremely difficult. Capturing information at multiple scales will assist in deciphering features, at global and local levels, to encompass lesions of variable size and texture. We introduce the Full-scale Deeply Supervised Attention Network (FuDSA-Net), for efficient segmentation of corona-infected lung areas in CT images. The model considers activation responses from all levels of the encoding path, encompassing multi-scalar features acquired at different levels of the network. This helps segment target regions (lesions) of varying shape, size and contrast. Incorporation of the entire gamut of multi-scalar characteristics into the novel attention mechanism helps prioritize the selection of activation responses and locations containing useful information. Determining robust and discriminatory features along the decoder path is facilitated with deep supervision. Connections in the decoder arm are remodeled to handle the issue of vanishing gradient. As observed from the experimental results, FuDSA-Net surpasses other state-of-the-art architectures; especially, when it comes to characterizing complicated geometries of the lesions. 
\end{abstract}
\section{Introduction}
\label{sec:intro}
The recent pandemic COVID-19 has disrupted life throughout the world. An effective characterization of its lesions on the lung, involving variable size and texture, holds promise in its early detection and improved prognosis. Artificial Intelligence (AI) can be effectively used to detect these abnormalities, and extract textural features corresponding to the virus-specific markers; thereby resulting in faster detection and analysis of the infection level. Comparing the segmentation performance of $U-$Net \cite{ronneberger2015u} and SegNet \cite{saood2021covid} demonstrates the efficacy of deep learning for identifying target COVID-19 lesions. Goncharov {\it et al.} \cite{goncharov2021ct} used a multi-task framework to simultaneously segment the lesions, followed by the classification of the patients based on the percentage of infected lung regions. However, it resulted in a lower Dice Score Coefficient (DSC). An attention mechanism was introduced \cite{oktay2018attention} as a way to highlight only relevant activations during training. This reduced computational resources wasted on irrelevant activations and improved network generalization. Deep supervision \cite{lee2015deeply}, in the hidden layers of a deep network, allowed learning of more discriminative features while overcoming the vanishing gradient problem.

The D2A $U-$Net \cite{zhao2021d2a} consists of attention modules incorporated into the basic $U-$Net architecture to boost performance, but yields a low recall score. This is indicative of the presence of False Negatives (FN), which is highly undesirable in the context of medical image segmentation. A pair of $U$-Nets were coupled \cite{xie2021duda} to initially extract the lung region and then emphasize the COVID-19 lesions. However the large number of parameters, in conjunction with limited data, could lead to overfitting.

Early-stage pathologies on CT scans (like GGOs) have low contrast and blurred edges, with varying shapes and sizes. It also becomes difficult to precisely segment a COVID-affected region of interest, due to possible interference from neighboring regions (such as the heart and bronchi). Tackling these challenges requires an advanced architecture to learn generalized patterns of the lesions exhibiting such variations in terms of shape, size and intensity. 

This research improves on the traditional encoder-decoder framework by leveraging features captured at various levels of the encoding path. These are fed to a novel attention module, which reweights the input feature map volume to place greater emphasis on the relevant activations pertaining to the target regions. Such use of multi-scalar features to modulate the weights of the attention mechanism helps capture both coarse, semantically-rich, global details along with the fine-grained, spatial information of the target locality. This enables learning of patterns corresponding to irregularly structured target lesions; mainly due to the involvement of multi-scalar information at all levels. This is termed as ``full-scale" in our nomenclature. Assignment of larger weights to the feature maps of interest within the entire input volume, along with highlighting relevant locations within them (while suppressing the rest) directs attention to the pertinent details in the target lesion region. This helps enhance the quality of segmentation. 
In the following sections, we describe the proposed model FuDSA-Net, followed by a demonstration of its effectiveness on publicly-available COVID-infection data from lung CT.

\section{Network Architecture and Attention Mechanism}
\label{sec:method}

Here we outline the  architecture of the proposed Full-scale Deeply Supervised Attention Network (FuDSA-Net) and describe its novel attention mechanism.

The standard encoder-decoder framework for medical image segmentation tasks \cite{ronneberger2015u} consist of skip connections at various levels. These combine the resultant feature map volume of the encoder with the input at the corresponding level of the decoder, to retain the  contextual spatial information  for improved localization of the target regions in the final segmented map. However, often inaccurate representations get carried over from the lower layers \cite{oktay2018attention}; particularly, as the stronger features are learned with the activation maps propagating deeper into the network. As a result, the segmentation is often inaccurate leading to poor recall scores. 

\begin{figure}[t]
    \centering
    \centerline{\includegraphics[width=9cm]{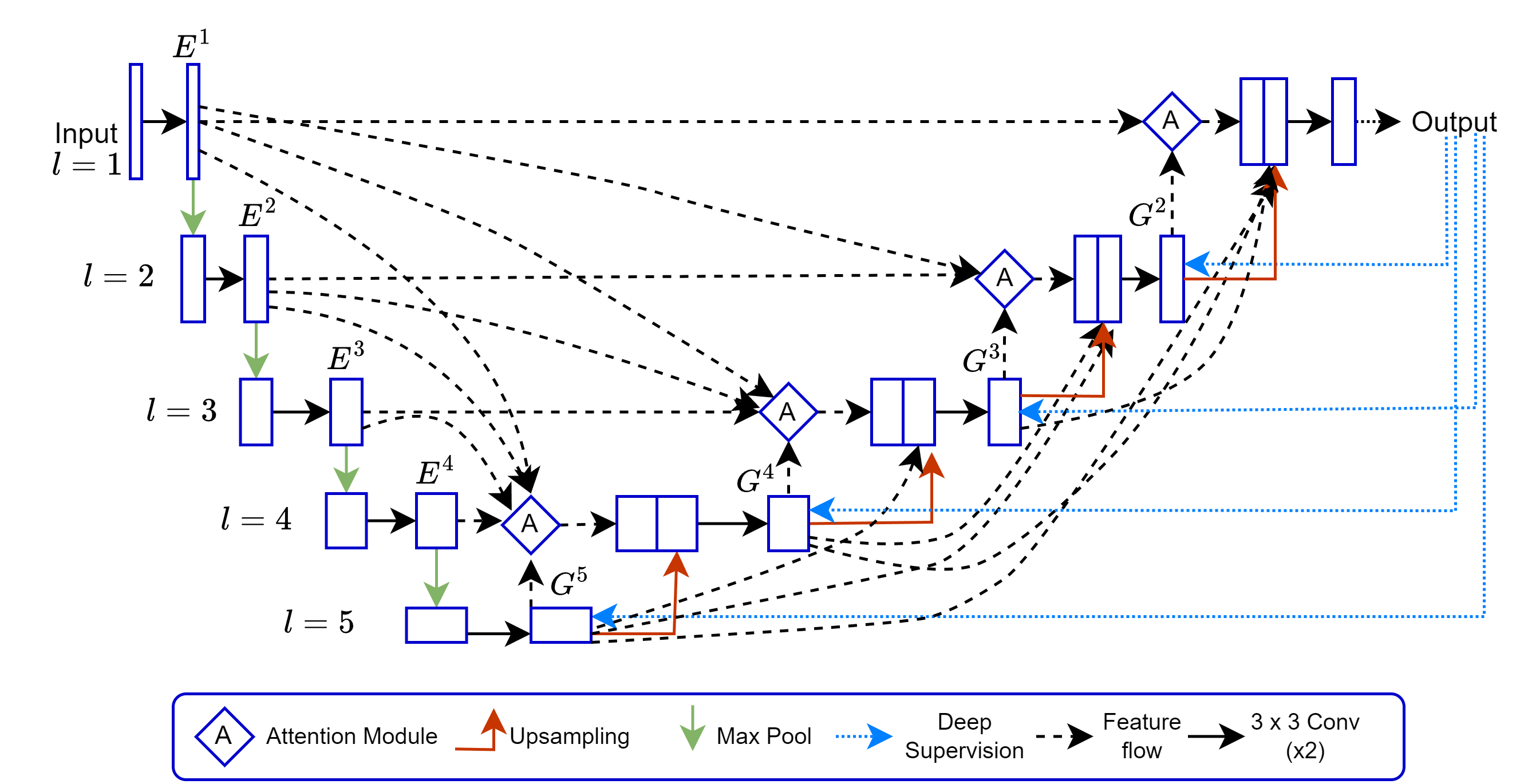}}
    \caption{FuDSA-Net for COVID-19 lesion segmentation.}
    \label{fig:fig1}
\end{figure}

To overcome this limitation, we introduce attention modules at various tiers to inhibit the activation(s) of the irrelevant region(s). This mechanism acts as a refinement for the encoder volume, before fusing with the decoder volume. The new attention scheme first locates the relevant feature maps across the entire input encoder volume, followed by the detection of important zones within them.  A combination of spatial and channel attention is employed for the purpose. Generating weights $\in [0,1]$, and multiplying them by the feature maps of the encoding path, dampen the activation in unimportant areas while amplifying the relevant response(s). Improved delineation of target lesions, encompassing varying shapes and sizes, is achieved through the incorporation of multi-scalar feature maps from all levels of the encoder while evaluating the necessary weights.

The proposed Full-scale Deeply Supervised Attention Network (FuDSA-Net), depicted in Fig. \ref{fig:fig1}, is built on the standard encoder-decoder framework of the $U$-Net. The skip connections are redesigned by incorporating the novel attention module at each level $l$. Additionally, the connections within the decoder arm are restructured to enable each layer to receive accumulated {\it knowledge} from all its preceding blocks. This mechanism promotes enhanced feature propagation along the entire decoding path, with  minimal loss of information. Deep supervision, applied throughout the decoder, helps learn discriminative patterns at all transitional levels. 

\begin{figure}[t]
    \centering
    \centerline{\includegraphics[width=9cm]{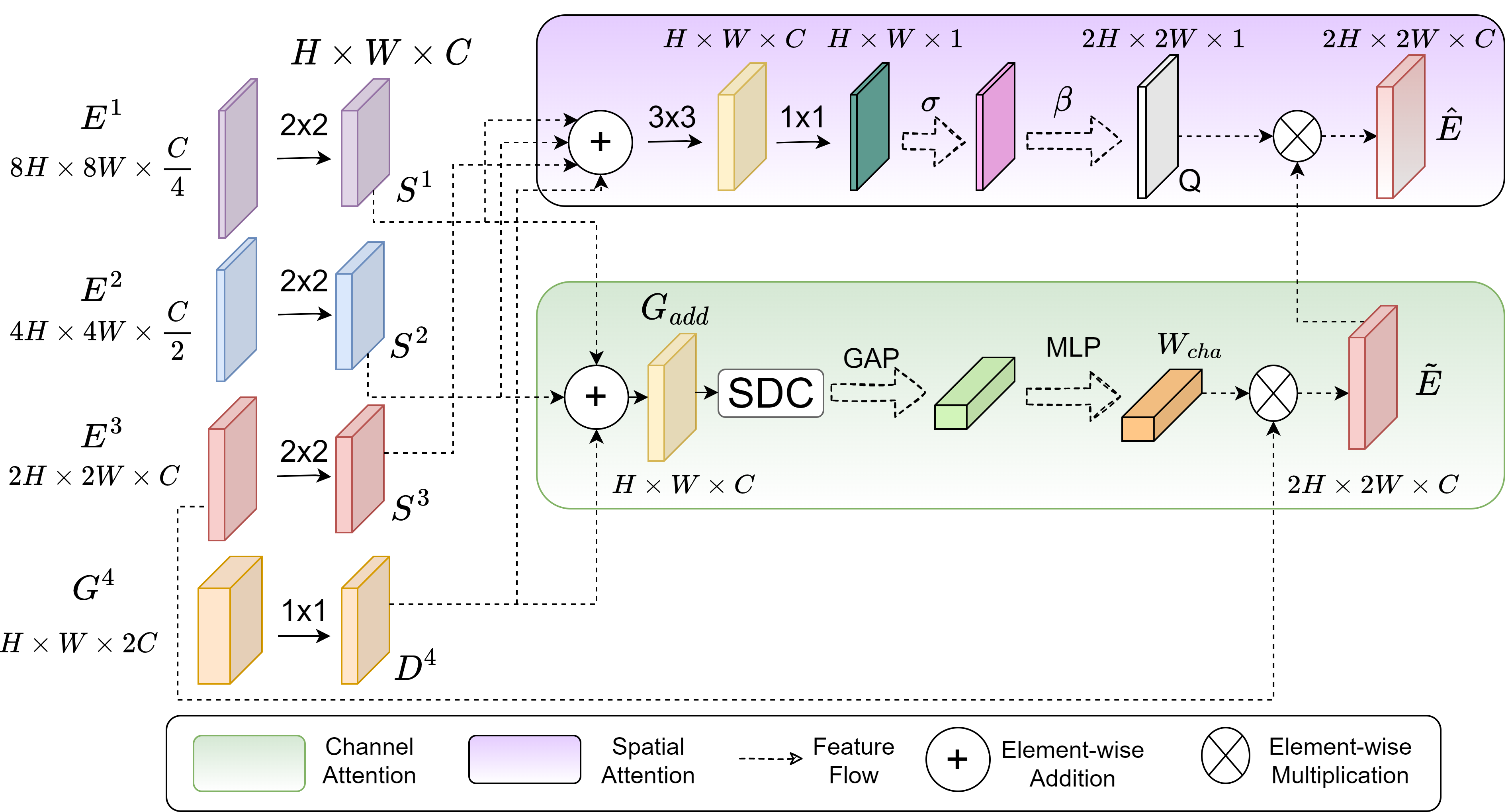}}
    \caption{Illustration of the working of  attention module, to produce a weighted encoder activation map volume $\hat{E}$ in the third level ($l=3$), of FuDSA-Net.}
    \label{fig:fig2}
\end{figure}

Fig. \ref{fig:fig2} illustrates how the attention module generates the recalibrated output volume, at the third level of the network, from the encoder volume $E^3$. The input to the attention module (at level $l$) is a collection of activation maps from {\it all} preceding levels of the encoder arm (up to $l$), encompassing {\it full} multi-scalar information expressed as $[E^1, E^2, \ldots, E^l]$ and the output volume of the decoder from level $(l+1)$ which is $G^{l+1} \in \mathbb{R}^{H \times W \times 2C}$. To maintain consistency across the input response maps from the encoder, over multiple scales, we perform a $2 \times 2$ convolution on the high-resolution maps to obtain the set of maps $[S^1, S^2, \ldots, S^l|S^i \in \mathbb{R}^{H \times W \times C}]$; thereby, reducing their spatial and channel dimensions until they are of the same dimensions as the lower-resolution maps. The decoder volume also undergoes a point-wise convolution to shrink its number of channels from $2C$ to $C$. The resultant decoder volume is denoted as $D^{l+1} \in \mathbb{R}^{H \times W \times C}$. 

The maps $[S^1, S^2, \ldots, S^{l-1}, D^{l+1}]$ are element-wise added ($\oplus$), along the channel attention branch, to produce an intermediate resultant volume $G_{add} \in \mathbb{R}^{H \times W \times C}$ expressed as 
\begin{equation}
    G_{add} = S^1 \oplus S^2 \oplus \ldots \oplus S^{l-1} \oplus D^{l+1}. 
\end{equation}
Next a block of several stacked dilated convolution (SDC) kernels, having varying dilation rates, is applied to $G_{add}$. This allows multi-scalar receptive fields, with the benefit of learning at various resolutions, by widening the kernel size without increasing the total number of parameters. Global Average Pooling (GAP) condenses the global information present in each map of the resultant volume into a tensor. This is passed through a multi-layer perceptron (MLP) with sigmoid activation function $\sigma$ and RELU to generate the final weight tensor $W_{cha} \in \mathbb{R}^{C}$, to be point-wise multiplied ($\otimes$) with the encoder map  $E^l \in \mathbb{R}^{2H \times 2W \times C}$ (at level $l$) to yield $\tilde{E} \in \mathbb{R}^{2H \times 2W \times C}$. We have
\begin{equation}
     W_{cha} =\sigma[MLP\{GAP(SDC \, \overline{G_{add}})\}],
\end{equation}
\begin{equation}
    \tilde{E} = W_{cha} \otimes E^l.
    \label{chan}
\end{equation}

Let us now consider the spatial attention component. Here the input maps $[S^1, S^2, \ldots, S^{l}, D^{l+1}]$ are pixel-wise added, followed by a $3 \times 3$ convolution $f_{3 \times 3}$ to learn features from the intermediate outputs. Next the resultant volume is convolved with $1 \times 1$ kernels $f_{1\times 1}$, followed by $\sigma$ activation and upsampling operation $\beta$  to generate the spatial attention map $Q \in \mathbb{R}^{2H \times 2W \times 1}$ as 
\begin{equation}
    Q = \beta [ \sigma \{f_{1 \times 1}(f_{3 \times 3} \,\,  \overline{S^1\oplus S^2\oplus \ldots \oplus S^{l}\oplus D^{l+1} } )\}].
    \label{spat}
\end{equation}
The final output $\hat{E} \in \mathbb{R}^{2H \times 2W \times C}$ of the attention module is generated, as a combination of the channel and spatial attention components of eqns. (\ref{chan}) and (\ref{spat}), as
\begin{equation}
    \hat{E} = Q \otimes \tilde{E}
\end{equation}

\section{Experimental Results}\label{results}

The model was trained using 2D CT slices from three publicly available datasets \cite{MedSeg2021}, \cite{jun2020covid} and \cite{morozov2020mosmeddata}. Preprocessing resized all slices to 512 × 512. The voxel intensities of all CT volumes, from three data sources, were  clipped to place them in the range [1000 HU, 170 HU] to filter out unnecessary details and noise. The inclusion criterion considered only those CT slices containing lesions. Combining such extracted (multi-source) slices, into a single set for the presentation, allowed the model a better exposure towards improved generalization; particularly, in  differentiating between various COVID-19 lesion structures and appearances corresponding to different severity levels. Intensity normalization was performed on the multi-source combined dataset. It was then randomly divided into 80\% for training and 20\% for evaluating the generalization performance.

FuDSA-Net is developed using Tensorflow and Keras in Python 3.9. All experiments are performed on a 12GB NVIDIA GeForce RTX 2080 Ti GPU. The focal Tversky loss function \cite{abraham2019novel} is used for training along with the Adam optimizer. The learning rate was set to $10^{-4}$ with an early stopping mechanism employed to prevent overfitting. The segmentation performance is determined by reporting the values of Dice Score Coefficient ($DSC$), Intersection over Union ($IoU$) and $Recall$ metrics.

\begin{table}[t]
    \centering
    \caption{Ablation study on FuDSA-Net}
    \begin{tabular}{c|c|c|c}
    \hline
     \textbf{Model}&\textbf{DSC}&\textbf{Recall}&\textbf{IoU}  \\
     \hline
     FuDSA-Net & \textbf{0.7924} & \textbf{0.8104} & \textbf{0.6681}\\
     \hline
     FuDSA-Net-I & 0.7424 & 0.7046 & 0.6074\\
     \hline
     FuDSA-Net-II & 0.7639 & 0.7246 & 0.6308\\
     \hline
     FuDSA-Net-III & 0.7905 & 0.7709 & 0.6650\\
     \hline
    \end{tabular}
    \label{tab:tab2}
\end{table}

To examine the role of the constituent components of FuDSA-Net, we did ablation studies involving three variants. These are (i) FuDSA-Net-I, with the spatial attention branch only; (ii) FuDSA-Net-II, with no deep supervision; and (iii) FuDSA-Net-III, excluding the residual connections between the various stages of the decoder. Experimental results for each of these variants are summarised in Table \ref{tab:tab2}. The best results are marked in bold in the table. It is observed that incorporating channel attention in FuDSA-Net significantly improves its performance over FuDSA-Net-1, as quantified by a sizeable  gain in $DSC$. This suggests that an enhanced attention mechanism is necessary in order to capture the complicated lesion regions of COVID-19. Improvement is also observed through the involvement of deep supervision as well as the incorporation of intra-stage connections in decoder arms.

\begin{table}[t]
    \centering
    \caption{Comparison of FuDSA-Net with baseline models}
    \begin{tabular}{c|c|c|c}
    \hline   \textbf{Model}&\textbf{DSC}&\textbf{Recall}&\textbf{IoU}  \\
     \hline
     FuDSA-Net & \textbf{0.7924} & \textbf{0.8104} & \textbf{0.6681}\\
     \hline
     $U$-Net & 0.6722 & 0.6515 & 0.5315\\
     \hline
     $U$-Net++ & 0.7167 & 0.6889 & 0.5774\\
     \hline
     Attention $U$-Net & 0.6480 & 0.6232 & 0.5071\\
     \hline
     Residual $U$-Net & 0.6476 & 0.6040 & 0.5094\\
     \hline
    \end{tabular}
    \label{tab:tab1}
\end{table}

A comparison of the proposed FuDSA-Net was also made with baseline architectures, like $U$-Net \cite{ronneberger2015u}, $U$-Net++  \cite{zhou2018unet++}, Attention $U$-Net  \cite{oktay2018attention}, and Residual $U$-Net \cite{khanna2020deep}, in terms of the different metrics. It is evident from Table \ref{tab:tab1}, that FuDSA-Net outperforms them by a significant margin of around 8\% in terms of $DSC$ (as compared to the scores by the second best-performing model, $U$-Net++). A $Recall$ value of $>80 \%$ signifies a fewer number of {\it False Negative} pixels in the generated output. A noticeable gain is also evident in terms of the $IoU$ value by our FuDSA-Net. 

\begin{figure}[t]

\begin{minipage}[b]{.48\linewidth}
  \centering  \centerline{\includegraphics[width=3.0cm,height=3.0cm]{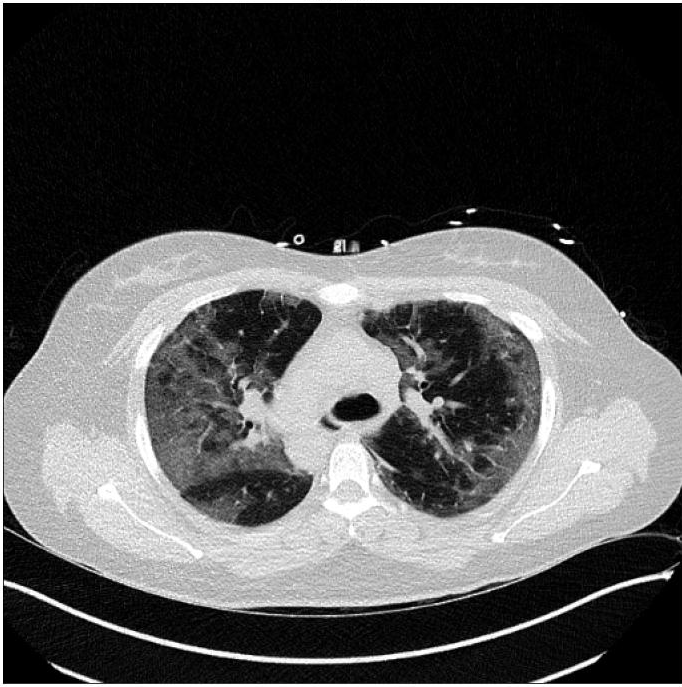}}
%  \vspace{1.5cm}
  \centerline{(a) Lung CT scan}\medskip
\end{minipage}
\hfill
%\hspace{1em}
\begin{minipage}[b]{0.48\linewidth}
  \centering
  \centerline{\includegraphics[width=3.0cm,height=3.0cm]{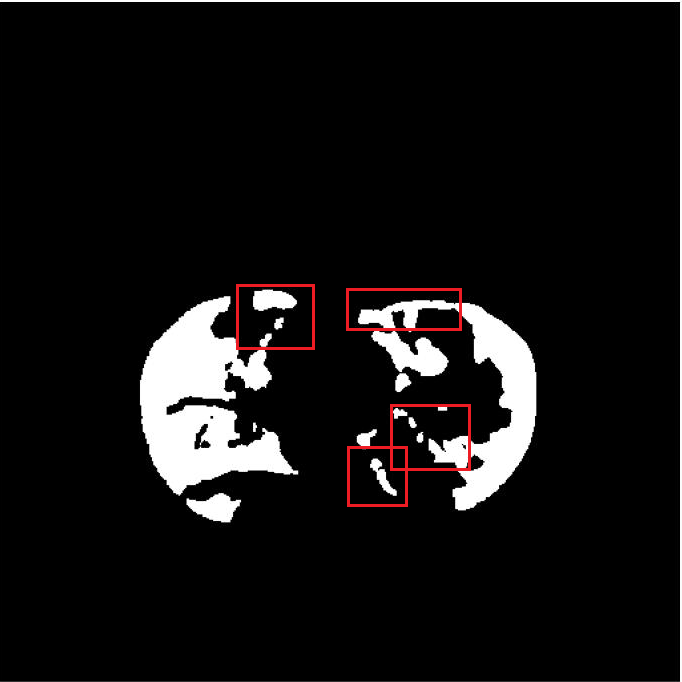}}
%  \vspace{1.5cm}
  \centerline{(b) Ground Truth}\medskip
\end{minipage}
\begin{minipage}[b]{.48\linewidth}
  \centering
 \centerline{\includegraphics[width=3.0cm,height=3.0cm]{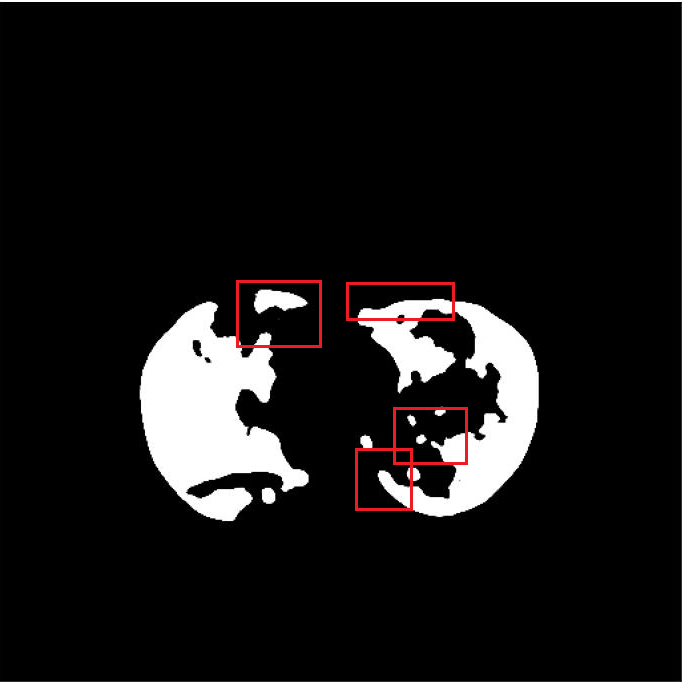}}
%  \vspace{1.5cm}
  \centerline{(c) Prediction by FuDSA-Net}\medskip
\end{minipage}
\hfill
%\hspace{1em}
\begin{minipage}[b]{0.48\linewidth}
  \centering
 \centerline{\includegraphics[width=3.0cm,height=3.0cm]{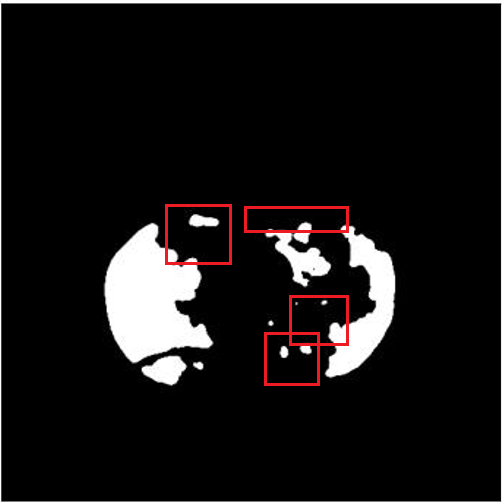}}
%  \vspace{1.5cm}
  \centerline{(d) Prediction by $U$-Net++}\medskip
\end{minipage}
\caption{(a) Sample CT scan of COVID-19 affected patient, with (b) corresponding ground truth, along with predictions made by (c) proposed FuDSA-Net, and (d) $U$-Net++. The red box highlights the comparison area in each case.}
\label{fig:res}
\end{figure}

The resultant segmentation map of FuDSA-Net is  observed to be relatively more accurate and closer to the corresponding ground truth, as illustrated in Fig. \ref{fig:res}. 

\section{Conclusion}\label{concl}

A novel deep learning architecture, FuDSA-Net, has been designed to effectively segment the COVID-19 lesions from lung CT scans. Multi-scalar features were acquired from all stages of the encoding path, for improved modeling of lesions having varying shapes, sizes and intensities. The attention mechanism was unique, incorporating both channel  and spatial attention for generating weights.  These were used for recalibrating the encoder map volume, prior to concatenation with the input volume at the decoder arm, to minimize unimportant activation from the input. Additional deep supervision enabled direct monitoring of intermediate layers in the decoding pathway. The vanishing gradient problem was avoided through the residual connections added all along the upsampling path. The experimental results demonstrated the superiority of FuDSA-Net, as compared to state-of-the-art methods, in identifying challenging target regions. 

\bibliographystyle{IEEEbib}
\bibliography{covid_biblio}

\begin{thebibliography}{10}

\bibitem{ronneberger2015u}
O.~Ronneberger, P.~Fischer, and {\it et al.},
\newblock ``{\it U}-{N}et: {C}onvolutional networks for biomedical image
  segmentation,''
\newblock in {\em Proccedings of the {I}nternational {C}onference on {M}edical
  {I}mage {C}omputing and {C}omputer {A}ssisted {I}ntervention, ({MICCAI})}.
  Springer, 2015, pp. 234\--241.

\bibitem{saood2021covid}
A.~Saood and I.~Hatem,
\newblock ``{COVID-19} lung {CT} image segmentation using deep learning
  methods: {$U$-N}et versus {S}eg{N}et,''
\newblock {\em {BMC} {M}edical {I}maging}, vol. 21, pp. 1\--10, 2021.

\bibitem{goncharov2021ct}
M.~Goncharov, M.~Pisov, and {\it et al.},
\newblock ``{CT-}based {COVID-19} triage: {D}eep multitask learning improves
  joint identification and severity quantification,''
\newblock {\em Medical {I}mage {A}nalysis}, vol. 71, pp. 102054, 2021.

\bibitem{oktay2018attention}
O.~Oktay, J.~Schlemper, and {\it et al.},
\newblock ``Attention {$U$-N}et: {L}earning where to look for the pancreas,''
\newblock in {\em Proceedings of the {M}edical {I}maging with {D}eep
  {L}earning}, 2018, pp. 1--10.

\bibitem{lee2015deeply}
Chen-Yu Lee, Saining Xie, and {\it et al.},
\newblock ``Deeply-supervised nets,''
\newblock in {\em Artificial {I}ntelligence and {S}tatistics}. PMLR, 2015, pp.
  562\--570.

\bibitem{zhao2021d2a}
X.~Zhao, P.~Zhang, and {\it et al.},
\newblock ``{D2A $U$-N}et: {A}utomatic segmentation of {COVID-19 CT} slices
  based on dual attention and hybrid dilated convolution,''
\newblock {\em Computers in {B}iology and {M}edicine}, vol. 135, pp. 104526,
  2021.

\bibitem{xie2021duda}
F.~Xie, Z.~Huang, and {\it et al.},
\newblock ``{DUDA-N}et: {A} double {U}-shaped dilated attention network for
  automatic infection area segmentation in {COVID-19} lung {CT} images,''
\newblock {\em {I}nternational {J}ournal of {C}omputer {A}ssisted {R}adiology
  and {S}urgery}, vol. 16, pp. 1425\--1434, 2021.

\bibitem{MedSeg2021}
MedSeg, H.~B. Jenssen, and T.~Sakinis,
\newblock ``Med{S}eg {COVID} {D}ataset 2,''
\newblock 2021.

\bibitem{jun2020covid}
J.~Ma, C.~Ge, and {\it et al.},
\newblock ``{COVID-19} {CT} {L}ung and {I}nfection {S}egmentation {D}ataset,''
  Apr. 2020.

\bibitem{morozov2020mosmeddata}
S.~P. Morozov, A.~E. Andreychenko, and {\it et al.},
\newblock ``M{OSMED} data: {D}ata set of 1110 chest {CT} scans performed during
  the {COVID-19} epidemic,''
\newblock {\em Digital {D}iagnostics}, vol. 1, pp. 49\--59, 2020.

\bibitem{abraham2019novel}
N.~Abraham and N.~M. Khan,
\newblock ``A novel focal {T}versky loss function with improved attention
  {U}-{N}et for lesion segmentation,''
\newblock in {\em Proceedings of {IEEE} 16th {I}nternational {S}ymposium on
  {B}iomedical {I}maging {(ISBI 2019)}}, 2019, pp. 683\--687.

\bibitem{zhou2018unet++}
Z.~Zhou, R.~Siddiquee, and {\it et al.},
\newblock ``U{N}et++: A nested {U}-{N}et architecture for medical image
  segmentation,''
\newblock in {\em Deep {L}earning in {M}edical {I}mage {A}nalysis and
  {M}ultimodal {L}earning for {C}linical {D}ecision {S}upport}. Springer, 2018,
  pp. 3\--11.

\bibitem{khanna2020deep}
A.~Khanna, N.~D. Londhe, and {\it et al.},
\newblock ``A deep {R}esidual {U}-{N}et convolutional neural network for
  automated lung segmentation in computed tomography images,''
\newblock {\em Biocybernetics and {B}iomedical {E}ngineering}, vol. 40, pp.
  1314--1327, 2020.

\end{thebibliography}

\end{document}